\documentclass[aps,prl,preprint,groupedaddress,nobibnotes,nofootinbib,byrevtex]{revtex4-1}
\usepackage[dvips]{graphicx}
\usepackage{amsmath,amssymb}
\usepackage{mathrsfs}

\begin{document}
\title{On transition of non-stationary waves}
\author{{Kenzo Ishikawa} and {Yutaka Tobita}}
\affiliation{{Department of Physics, Faculty of Science, Hokkaido University, Sapporo 060-0810, Japan}
}
\date{\today}
\begin{abstract}
{The  probability of the events that the final states are detected with or
 interact with the nucleus in a finite time interval T was found to 
be, 
$P=\text T \Gamma_0 +P^{(d)}$.  $\Gamma_0$ is computed with Fermi's
 golden rule, and does not depend on the nuclear wave functions. 
 $P^{(d)}$ is not given by  Fermi's
 golden rule, and  depends on the nuclear wave functions. 
  In the electron mode of pion decays, $\Gamma_0$ is proportional 
to $m_e^2$ but $P^{(d)}$ for the event that the neutrino is detected 
is proportional to  $m_{\nu}^{-2}$. $P^{(d)}$ does not hold  the
 helicity suppression satisfied in   $\Gamma_0$  and   is 
inevitable  in  non-stationary quantum phenomena.}
\end{abstract}
\preprint{EPHOU-13-011}
\maketitle
\section{Transitions of non-stationary waves } 
The transition rates   are  computed using the  initial 
and final states at the asymptotic regions, which  satisfy 
boundary conditions of the infinite time-interval $\text T=\infty$.
Stationary phenomena are studied  with methods for  stationary waves.
  Now the non-stationary phenomena  have non-trivial T-dependence    
and  are studied with T-dependent transition probabilities, which can
not be given from the stationary method.  They are 
computed 
with a method of non-stationary waves, which satisfy boundary conditions
different from those of stationary waves. The method  is applied  to a
neutrino in pion decays in the 
present paper. We show, using
the non-stationary method, that  the 
transition probability  at large T has a constant term in addition to a T-linear term
\cite {ishikawa-tobita}, and is expressed as 
\begin{eqnarray}
P=\text T \Gamma_0 +P^{(d)}. \label{probability0}
\end{eqnarray}
The rate $\Gamma_0$ is computed with  Fermi's golden rule 
\cite{Dirac,Schiff-golden} and satisfies known
properties.   Now $P^{(d)}$ is not computed with the stationary waves 
and   was obtained with the non-stationary method in Ref. \cite {ishikawa-tobita}.  
 $P^{(d)}$ possesses unique properties that $\Gamma_0$ does not possess,
 and  is important in various situations, especially in  processes of   
$\Gamma_0 = 0$, $\text T \approx 0$. An example is presented.

In  a microscopic system  
described  
by the Lorentz invariant   Lagrangian,     
\begin{align}
\mathcal{L} =\mathcal{L}_0+\mathcal{L}_{int},
\end{align}
the state vectors  can be defined in Poincar\'e covariant manner.  They 
are expressed by a many body wave function $|\Psi
 \rangle$ that  follows a Schr\"{o}dinger equation
\begin{eqnarray}
i\hbar{\partial \over \partial t}|\Psi\rangle=(H_0+H_{int})|\Psi \rangle,
\label{Schroedinger}
\end{eqnarray} 
where the free part, $H_0$, and the interaction part, $H_{int}$, are derived
 from the previous Lagrangian.  The transition processes are studied 
with Eq. ($\ref{Schroedinger}$) and classified to various cases 
 depending on a situation  of the system.  

The transitions of  lower excited states  are studied with
 S-matrix  $S[\infty]$ defined with an initial state at $t=-\infty$ 
and  a final state at
$t=\infty$. When the former states  and the
latter  states are  non-interacting each other and the overlap between 
them are negligible,
 the decay rates and cross section are considered  to those of 
isolated particles in vacuum.
 The  asymptotic quantities 
at $\text{T}=\infty$  thus  computed with $S[\infty]$ are  Poincar\'e
 invariant  \cite{Goldberger,newton,taylor}.
The transition probability is proportional to T,  
\begin{eqnarray}
P=\text T \Gamma_0 \label{T-linear probability},
\end{eqnarray}
and $\Gamma_0$ is computed with Fermi's golden rule.
 Waves of small spatial
 extensions  follow Eq. $(\ref{T-linear probability})$. Their successive  
reactions occur
 separately,  and 
the probability 
becomes in-coherent sum of those of each  process.  
This region has been studied well and is called 
particle zone, here.

Transition of  non-stationary waves reveals different probability. At a
finite T, an overlap of the 
in-coming or out-going waves  is not ignored  and  modifies the 
probabilities from those of the
 asymptotic region. The probability of successive processes   becomes 
non-factorized  to each  process and  the probability
of the  event that the final state interacts with others  shows  a
 different behavior  from the particle zone. 
 The  detector in experiments is composed of many atoms and the 
outgoing particle in the event  interacts with them.    These  
scatterings  occur at a finite T,
instead of the infinite T. Accordingly, an S-matrix of the finite-time 
interval T, denoted as $S[\text T]$, was introduced \cite{ishikawa-tobita}.   $S[\text T]$ is defined  in such way
 that satisfies the boundary condition at T  with  M{\o}ller operators 
at a finite T,
 $\Omega_{\pm}(\text{T})$, as $S[\text T]=\Omega_{-}^{\dagger}(\text T)\Omega_{+}(\text
T)$.  $\Omega_{\pm}(\text{T})$  are expressed  by a free Hamiltonian  $H_0$ 
and  a  total Hamiltonian  $H$ by  $\Omega_{\pm}(\text T)=\displaystyle\lim_{t
 \rightarrow \mp \text T/2}e^{iHt}e^{-iH_0t}$.  From this expression,     
$S[\text T]$ satisfies
\begin{align}
&[S[\text T],H_0]=  i\left\{\frac{\partial}{\partial
	\text{T}}\Omega^{\dagger}_{-}(\text{T})\right\}\Omega_{+}(\text{T})-
	i\Omega^{\dagger}_{-}(\text{T}){\partial \over \partial \text{T}}
	\Omega_{+}(\text{T}).
\label{commutation-relation}
\end{align} 

Due to Eq. $(\ref{commutation-relation})$,  a  matrix
element of $S[\text T]$ between two eigenstates of $H_0$, $|\alpha \rangle$ 
and  $|\beta \rangle $ of eigenvalues $E_{\alpha}$ and $E_{\beta}$, 
is decomposed into  two components
\begin{eqnarray}
\langle \beta |S[\text{T} ]| \alpha \rangle=\langle \beta
 |S^{(n)}[\text{T} ]| \alpha \rangle+ \langle \beta |S^{(d)}[\text{T} ]|
 \alpha \rangle,
\label{matrix-element}
\end{eqnarray}
where $\langle \beta |S^{(n)}|\alpha \rangle$  becomes finite for 
$E_{\beta}=E_{\alpha}$ and
$\langle \beta |S^{(d)}|\alpha \rangle$  becomes finite for  $E_{\beta} \neq E_{\alpha}$.
 The  first term is equivalent to the asymptotic value and 
the second term  is  added to it. Thus the probability 
has a correction  to  Fermi's golden rule. The fact that the 
correction is derived from the kinetic-energy non-conserving term and  
may modify  dynamical properties  was  known to Peierls and 
Landau \cite{peierls}, and Landau concluded
that the correction was negligible for processes of small energy
transfer. We  showed  that the corrections are 
in fact important for processes of large energy transfer in the previous 
work \cite{ishikawa-tobita}, and the probability  has a $T$-linear and
constant, Eq. $(\ref{probability0})$.
$\Gamma_0$ is computed with $S[\infty]$ and has been studied literature,
 but the $P^{(d)}$ can not be computed  with $S[\infty]$ but by $S[\text T]$.
 
The states of continuous 
spectrum of kinetic-energy  couple with  $S^{(d)}[\text T]$.  
Among the infinite number of states of $|\beta \rangle $ of $E_{\beta}
  \neq E_{\alpha}$, certain states satisfy
boundary conditions at $t=\pm {\text T}/2$. They  are expressed  
for the
 scalar field $\phi(x)$ with 
 field operators  \cite
 {LSZ,Low,ishikawa-tobita} 
\begin{eqnarray}
& &\lim_{t \rightarrow - \text T/2}\langle\alpha| \phi^f |\beta
 \rangle=  \langle \alpha| \phi_{in}^f |\beta \rangle,\\
& &\lim_{t \rightarrow + \text T/2}\langle \phi^f |0 \rangle=  \langle
 \alpha| \phi_{out}^f |\beta \rangle,
\end{eqnarray}
where $\phi_{in}(x)$ and $\phi_{out}(x)$ satisfy the free wave equation,
and $\phi^f,\ \phi_{in}^f$ and $\phi_{out}^f$ are the expansion
coefficient of  $\phi(x),\ \phi_{in}(x)$ and $\phi_{out}(x)$, with the
normalized wave functions $f(x)$ of the form
\begin{eqnarray}
\phi^f(t)=i\int d^3 x f^{*}({\vec x},t)\overleftrightarrow{\partial_0}  \phi({\vec x},t).
\label{wave-packet-expansion}
\end{eqnarray}
For events that the neutrino is detected with the nucleon in the
 detector, the probability amplitude is expressed with the nucleon wave 
function in nucleus. Hence the nucleon wave function is used for $f(x)$.
 $S^{(d)}[\text T]$ thus defined depends on the base functions $f(x)$,
 and is appropriate to write as  $S^{(d)}[\text T;f]$. 
The neutrino  in the final state is  expressed by the small wave
 function despite of its large mean free path.
Accordingly, the probability of the events   is 
expressed by this  normalized wave function, called wave packet. 
Wave
  packets that satisfy  free wave equations and are localized in space 
are important  for rigorously defining scattering amplitude  
\cite {LSZ,Low}.

At $\text T \rightarrow \infty$, the right-hand side
of Eq.\,$(\ref{commutation-relation})$   and  the second term of
Eq.\,$(\ref{matrix-element})$ vanish, 
hence the energy defined by $H_0$ is  conserved. Conversely, all the
states $|\beta \rangle$ of $E_{\beta}=E_{\alpha}$ contribute and
$S[\infty]$ is uniquely defined. At a finite T, $S^{(n)}[\text T]$
satisfies the conservation of kinetic energy and is uniquely defined.
$S^{(d)}[\text T;f]$, on the other hand,  does not satisfy the conservation
of the kinetic energy  and depends
on $f$. Furthermore,  $S^{(d)}[\text T;f]$ is not invariant 
under the Poincar\'e transformation
  defined by $\mathcal{L}_0$. 
The state $|\beta \rangle $ of $E_{\beta}$ is orthogonal to 
$|\alpha \rangle $ of $E_{\alpha} \neq E_{\beta}$  and the cross term 
of the  first  and second terms of
Eq.\,$(\ref{matrix-element})$  in a square of the modulus vanishes. 
Consequently  the  finite-size 
correction to the probability becomes   positive
semi-definite, and the probability $P(\text T)$ is  larger than
$P(\infty)$. Unitarity $S[\text T]S^{\dagger}[\text T] =1$ is
satisfied and ensures the conservation of probability.  For
$S[\infty]$, the states have constant kinetic energy, and reveal the
particle properties.  The states of 
continuous kinetic energy are different and are  waves. 
It will be shown that they  reveal features of wave
  properties.  
\section{Forbidden process}
For a process $\langle \beta |S^{(n)}[\text T]|\alpha \rangle=0$, the
rate vanishes, $\Gamma_0=0$, and $P$
 is not proportional to  T but is  constant, $P^{(d)}$. $P^{(d)}$    is 
computed from $\langle \beta |S^{(d)}[\text
  T;f]|\alpha \rangle$. Thus
states   $|\beta \rangle$ of   $E_{\beta} \rightarrow  \infty $ 
that satisfy the boundary conditions contribute    to $P^{(d)}$. 
 These waves
 propagate  with the speed of light,
hence  necessarily give a finite contribution 
 to the probability of the event that a  light particle  is 
detected.

 Probabilities at finite T using   $S[\infty]$ were studied without wave
 packets  in
 Refs. \cite{khalfin,winter,ekstein,gaemer,peres,maiani-testa} and with 
wave packets    in Refs. \cite{Kayser,Giunti,Nussinov,Kiers,Stodolsky,Lipkin,Akhmedov,Asahara}.
The probabilities   from $S[\infty]$ do not show  the finite-size
corrections, because
they were  the asymptotic values. The probabilities from  $S[\text
T]$ are different, and show T-dependence.  For the event that the
 neutrino 
is detected at T,  
$S[\text T]$ with the  wave packet were studied in Refs.
\cite{Ishikawa-Shimomura,Ishikawa-Tobita-ptp,Ishikawa-Tobita,Ishikawa-Tobita-prl}. 
  The T-dependent probabilities  are derived from $S^{(d)}[\text T;f]$,
 and depend 
on $f$.  The wave packets can not be replaced 
with plane waves in $S[\text T]$. 


{\Large $\pi \rightarrow \nu_e+e$}  

 The probabilities of the events
 that the electron neutrino from the
   decay of pion interacts with the nearby nucleus  
   at a finite distance  satisfies $\Gamma_0 \approx 0,\ P^{(d)} \neq 0$ 
and is studied in this section. The system is
   descried by
\begin{align}
&\mathcal{L}=\mathcal{L}_0+\mathcal{L}_{int},\nonumber\\
&\mathcal{L}_0={\partial_{\mu}}\varphi^{*}{\partial^{\mu}}\varphi-m_{\pi}^2 \varphi^{*}\varphi+\bar l (\gamma \cdot 
 p-m_l)l+\bar\nu(\gamma \cdot p-m_{\nu})\nu \nonumber, \\
&\mathcal{L}_{int}= g J_{hadron}^{V-A}\times J_{lepton}^{V-A},\ g= G_F/{\sqrt
 2},
\end{align}
where $G_F$ is the Fermi coupling constant and $J_i^{V-A}$ is $V-A$
current.
The  amplitude  for a neutrino of an average momentum ${\vec
p}_{\nu}$ at $\vec{\text{X}}_{\nu}$ in the decay of a pion prepared at
$t=\text{T}_{\pi}$ of a momentum ${\vec p}_{\pi}$ is 
 expressed as
 $\mathcal{M}=\int d^4x \, \langle {l},{\nu}
 |H_{w}(x)| \pi \rangle$, where   a lepton $l$ has a momentum ${\vec p}_{l}$ 
and a  neutrino  is expressed by a wave packet.    These states  are
expressed as    
$
|\pi \rangle=   | {\vec p}_{\pi},\text{T}_{\pi}  \rangle,\ 
|l ,\nu \rangle=   |\mu,{\vec p}_{l};\nu,{\vec p}_{\nu},\vec{\text{X}}_{\nu},\text{T}_{\nu}          \rangle$.
  $\mathcal{M}$ is
written 
 with the hadronic matrix element of $V-A$ current and  Dirac spinors 
\begin{align}
\label{amplitude}
&\mathcal{M} = \int d^4xd{\vec k}_{\nu}
\,N_1\langle 0 |J_{V-A}^{\mu}(0)|\pi \rangle 
\bar{u}({\vec p}_l)\gamma_{\mu} (1 - \gamma_5)\nu({\vec k}_{\nu})\nonumber\\
&\times \exp{\left[{-i{(p_{\pi}-p_l) \cdot x /\hbar} + 
i{k_\nu\cdot(x - \text{X}_\nu)/\hbar}
 -\frac{\sigma_{\nu}}{2}({\vec k}_{\nu}-{\vec p}_{\nu})^2}\right]},
\end{align}
where 
$N_1=ig \left({\sigma_\nu/\pi}\right)^{\frac{4}{3}}\left({m_l m_{\nu}}/{
(2\pi)^3E_{\pi}V  E_l E_{\nu}}\right)^{\frac{1}{2}}$ and $\langle
0|J_{V-A}^{\mu}(0)|\pi \rangle=if_{\pi}p_{\pi}^{\mu}$.  
 The time $t$ is
 integrated over the region $\text{T}_{\pi} \leq t \leq \text{T}_{\nu}$,
in the region $\text{T}_{\nu}- \text{T}_{\pi} \ll \tau_{\pi}$, where
 $\tau_\pi$   is the pion life. 
${\sigma_{\nu}}$ is the size
of the nucleon wave function that the neutrino interacts with and is  
estimated from   the size of a nucleus.
For the sake of simplicity, we use the Gaussian form of the wave packet
 for the function $f(x)$ of Eq. $(\ref{wave-packet-expansion})$. 
The  result for   the finite-size correction is the same in general 
wave packets.

The  total probability  is computed from the amplitude as
\cite{Ishikawa-Shimomura},
\begin{eqnarray}
P=\int d\vec{\text{X}}_{\nu} {d{\vec p}_{\nu} \over (2\pi)^3} 
  \frac{d{\vec
 p}_l}{(2\pi)^3} \sum_{s_1,s_2}|\mathcal M|^2,
\label{total-probability}
\end{eqnarray}
 where the unmeasured momentum of the final state  is  integrated over
the whole positive energy region and that of the measured momentum and 
the position  is integrated in the inside  of the detector. 
 Hereafter the natural unit,
 $c=\hbar=1$, is taken in majority of places, but $c$ and $\hbar$ are
written explicitly when it is necessary.

 Equation \,$(\ref{total-probability})$ was computed in Ref.
 \cite{ishikawa-tobita}.
Integrating over the neutrino's coordinate $\vec{\text{X}}_{\nu}$,
we obtain  the total volume, which   is canceled by the factor 
$V^{-1}$ from  the normalization of the initial pion state.
The total
probability is then expressed as the sum of  the standard term, $\Gamma_0$, 
and the new term proportional to $\tilde g(\omega_\nu\text{T})$:
\begin{align}
\label{probability-3}
& P=\text T\Gamma_0+P^{(d)} \\
&\Gamma_0=\tilde N_4\int \frac{d^3 p_{\nu}}{(2\pi)^3}
\frac{p_{\pi}\! \cdot\! p_{\nu}(m_{\pi}^2-2p_{\pi}\! \cdot\! p_{\nu}) }{E_\nu}
    G_0 ,\nonumber \\
& P^{(d)}=\tilde N_4\int \frac{d^3 p_{\nu}}{(2\pi)^3}
\frac{p_{\pi}\! \cdot\! p_{\nu}(m_{\pi}^2-2p_{\pi}\! \cdot\! p_{\nu}) }{E_\nu}
   \text T \tilde g(\omega_{\nu}\text{T}),\nonumber
\end{align}
where $\tilde N_4 = 8g^2 f_{\pi}^2\sigma_\nu/E_{\pi}$ and $\text{L} = c\text{T}$ is the
length of the decay region, and  
 $\omega_{\nu}={m_{\nu}^2 \over 2E_{\nu}}$.   $G_0$  comes   from term of
$p_{\pi}\approx p_e+p_{\nu_e}$,  and 
\begin{eqnarray}
(m_{\pi}^2-2p_{\pi}\! \cdot\! p_{\nu}) =m_e^2,
\end{eqnarray}
 which  vanishes for $m_e=0$. The $\Gamma_0$ becomes independent of
 $\sigma_{\nu}$ and is proportional to the square  of  charged lepton mass. The ratio
 between the electron and muon is   $({m_e
 \over m_{\mu}})^2=10^{-4}$  and  $\Gamma_0$ for the electron 
is negligibly small, which is known as 
the helicity suppression.

 $\tilde
g(\omega_{\nu}\text{T})$  comes from the kinetic-energy non-conserving
term and  satisfies
$\tilde g(0)=\pi,\ {\partial \over \partial \text{T}}\tilde g(\omega_{\nu}\text{T})|_{\text{T}=0}
 = -\omega_{\nu}$ and  $\tilde g(\omega_{\nu}\text T)=\frac{2}{
 \omega_{\nu} \text T},\text{ for } \omega_{\nu}\text T \rightarrow \infty$. 
  In small T, $\tilde g(\omega \text T)=\tilde g(0)=\pi$, and  $P^{(d)}$ is
  proportional to T, and the probability from $\text T \Gamma_0$  is
ignorable. Hence 
\begin{align}
\label{probability-3'}
P= \text {T} \frac{8\pi g^2f_{\pi}^2\sigma_{\nu}  }{E_{\pi}} \int \frac{d^3 p_{\nu}}{(2\pi)^3}
\frac{p_{\pi}\! \cdot\! p_{\nu}(m_{\pi}^2-2p_{\pi}\! \cdot\! p_{\nu}) }{E_\nu}\theta(m_{\pi}^2-m_e^2-2p_{\pi}\cdot p_{\nu}).
\end{align}
 The  kinetic-energy non-conserving term 
 from the momentum region 
\begin{eqnarray}
& &p_{\pi}-( p_e+p_{\nu_e}) \neq 0, \\
& &(m_{\pi}^2-2p_{\pi}\! \cdot\! p_{\nu})
 \gg m_e^2 \nonumber
\end{eqnarray}    
gives $P^{(d)}$. Consequently the integral does not hold the helicity 
suppression, and  
$P/{\text T}|_{\text T \rightarrow 0}$ is sizable of around $20$ percent
of the rate of muon-mode. 
At a large macroscopic T,  $\tilde
g(\omega_{\nu} \text T) \approx 4E_{\nu}/({m_{\nu}^2\text{T}}) $, and we have,
\begin{align}
\label{probability-4}
P=\frac{16g^2 f_{\pi}^2\sigma_{\nu} \hbar}{E_{\pi} m_{\nu}^2c^3} \int \frac{d^3
 p_{\nu}}{(2\pi)^3}
{p_{\pi}\! \cdot\! p_{\nu}(m_{\pi}^2-2p_{\pi}\!
 \cdot\! p_{\nu}) \theta (m_{\pi}^2-m_e^2-2p_{\pi}\cdot p_{\nu})}.
\end{align}
$P$  at a large T is independent of T, which reveals the wave nature. The
  integral in Eq. $(\ref{probability-3'})$ is Lorentz invariant but that 
in Eq. $(\ref{probability-4})$ is 
  Lorentz non-invariant  even in high-energy region. Thus  $P$ is  
governed by $P^{(d)}$,  Eq. ($\ref{probability-3'}$) or
  Eq. ($\ref{probability-4}$) and has sizable magnitudes, which   increase
  with  $\sigma_{\nu}$ because the number of  kinetic-energy non-conserving
states increases.   Furthermore, the constant $P^{(d)}$ of
  Eq. ($\ref{probability-4}$) is inversely proportional to the square of
  the neutrino mass. $P^{(d)}$ is consistent with existing data on
  neutrino experiments \cite{particle-data} if $\sigma_{\nu}$ is the
  size of nucleus, and  a future precision 
measurement of  $P^{(d)}$ at $\text T \approx 0$
  and at a macroscopic  T will be able to supply the precise value of 
absolute neutrino  mass.
Implications of $P^{(d)}$ and other  processes of $\Gamma_0=0, P^{(d)}
  \neq 0$ will be studied  in a subsequent paper.

\begin{acknowledgments}
 This work was partially supported by a 
Grant-in-Aid for Scientific Research (Grant No. 24340043). The authors  
thank Dr. Nishikawa, Dr. Kobayashi, and Dr. Maruyama for useful discussions on 
the near detector of the T2K experiment and Dr. Asai, Dr. Kobayashi,
Dr. Mori, and Dr. Yamada
for useful discussions on interferences. 
\end{acknowledgments}

\appendix

\end{document}